\documentclass[]{spie}  

\usepackage{amsmath,amsfonts,amssymb}
\usepackage{graphicx}
\usepackage[colorlinks=true, allcolors=blue]{hyperref}
\usepackage{astro_bib_macro}

\usepackage{booktabs}

\usepackage{listings}

\title{Wavefront error tolerancing for direct imaging of exo-Earths with a large segmented telescope in space}

\author{Iva Laginja\supit{a}\supit{b}, Lucie Leboulleux\supit{c}, Laurent Pueyo\supit{a}, R\'{e}mi Soummer\supit{a}, Jean-Fran\c{c}ois Sauvage\supit{b}\supit{e}, Laurent Mugnier\supit{b}, Laura E. Coyle\supit{d}, J. Scott Knight\supit{d}, Kathryn St.Laurent\supit{a}, Emiel H. Por\supit{f}, James Noss\supit{a}
\skiplinehalf
\supit{a} Space Telescope Science Institute, 3700 San Martin Drive, Baltimore, MD 21218, USA\\
\supit{b} Office National d'\'{E}tudes et de Recherches A\'{e}rospatiales, 
92320 Ch\^{a}tillon, France\\
\supit{c} LESIA, Observatoire de Paris, Universit\'{e} PSL, CNRS, 
92195 Meudon, France\\
\supit{d} Ball Aerospace \& Technologies Corp., 1600 Commerce Dr., Boulder, CO 80303\\
\supit{e} Aix Marseille Universit\'{e}, CNRS, LAM 
UMR 7326, 13388 Marseille, France\\
\supit{f} Leiden Observatory, Leiden University, P.O. Box 9513, 2300 RA Leiden, The Netherlands
}

\authorinfo{Further author information, send correspondence to Iva Laginja: E-mail: ilaginja@stsci.edu, Telephone: 1 667 218 6530}

\begin{document} 
\maketitle

\begin{abstract}
Direct imaging of exo-Earths and search for life is one of the most exciting and challenging objectives for future space observatories.
Segmented apertures in space will be required to reach the needed large diameters beyond the capabilities of current or planned launch vehicles. 
These apertures present additional challenges for high-contrast coronagraphy, not only in terms of static phasing but also in terms of their stability. 
The Pair-based Analytical model for Segmented Telescope Imaging from Space (PASTIS) was developed to model the effects of segment-level optical aberrations on the final image contrast. In this paper, we extend the original PASTIS propagation model from a purely analytical to a semi-analytical method, in which we substitute the use of analytical images with numerically simulated images. The inversion of this model yields a set of orthonormal modes that can be used to determine segment-level wavefront tolerances. We present results in the case of segment-level piston error applied to the baseline coronagraph design of LUVOIR A, with  minimum and maximum wavefront error constraint between 56 pm and 290 pm per segment. The analysis is readily generalizable to other segment-level aberrations modes, and can also be expanded to establish stability tolerances for these missions. 
\end{abstract}


\section{INTRODUCTION}
\label{sec:introduction}

Imaging Earth-like planets and searching for biomarkers is one of the key science objectives in space astronomy for the next decade. The capability to reach this ambitious goals is a steep function of the primary mirror diameter, which drives the missions designs toward large apertures \cite{2014ApJ...795..122S,stark2015ApJ}. The Large UV Optical Infrared Surveyor (LUVOIR)\cite{luvoir-interim-report} and the Habitable Exoplanet Observatory (HabEx)\cite{habex-interim-report} are being studied toward this goal as part of a series of mission concept studies . The LUVOIR study has two point-design cases (LUVOIR-A and LUVOIR-B), respectively 15 m and 8 m in diameters, each with a suite of scientific instruments that include coronagraphs. In both LUVOIR point-designs, the primary science objective is the direct detection and spectral characterization of habitable Earth-like planets and the search for life. Both have in common that their primary mirror is segmented, much like the James Webb Space Telescope (JWST) to allow for large light collecting areas beyond the capabilities of plausible monolithic mirrors given current or planned launch capabilities. However, telescope segmentation introduces additional sources of wavefront errors from segment cophasing, as well as diffraction effects from segment gaps\cite{juanola2019spie}. Given the required star attenuation levels of $10^{-10}$ to $10^{-11}$ to image exo-Earths, the observatory not only has to reach these contrasts, but also has to maintain them over appropriate observation time scales. 

These high-contrast goals with segmented apertures impose severe requirements both on the static wavefront control, but also the overall stability of the mechanical structures of the telescope. Conceptually, this problem can be divided into different spatial regimes and temporal timescales\cite{pueyo2019spie, coyle2019spie}. Between low spatial scales that mostly arise from global telescope misalignments and high spatial scales that come primarily from static polishing errors, the mid spatial regime encompasses modes caused by primary mirror misalignments and local aberrations on the individual segments. It is these mid-spatial frequency scales that we study in the present work, building on the Pair-based Analytical model for Segmented Telescope Imaging from Space (PASTIS) that was developed for high-contrast coronagraphy with segmented apertures\cite{leboulleux2018, leboulleux2018spie}.

The goal of PASTIS is to model the effects of segment-level optical aberrations on the final image contrast, and therefore provide a framework to establish wavefront and stability requirements. For a given telescope geometry and coronagraph design, PASTIS provides a framework to identify dominant mid-spatial frequency modes resulting from the primary mirror segmentation. The sensitivity of the dark-zone contrast can be established in relation to these so-called PASTIS modes, and the dynamic tolerances further defined by including considerations of the sensing and control system (in particular their timescales and efficiencies) for a given target contrast. In this paper, we focus the application of PASTIS on establishing static wavefront error tolerances, and we leave the derivation of dynamic drift rates to future work. The original PASTIS model is a fully analytical framework that constructs a matrix that can then be used for contrast calculations and, by its inversion, wavefront error tolerancing. The work presented here extends PASTIS to a semi-analytical matrix generation, which we demonstrate on the LUVOIR-A telescope with an Apodized Pupil Lyot Coronagraph (APLC)\cite{2005ApJ...618L.161S,ndiaye2015,ndiaye2016,zimmerman2016jatis,2016SPIE.9904E..1YZ,St.Laurent2018spie}.

In Sec. \ref{sec:top-level-pastis} we recall the PASTIS model applied to perfect and real coronagraphs and introduce the extension to the semi-analytical matrix calculation. Sec. \ref{sec:tolerancing} describes the calculation of mode- and segment-based wavefront error tolerances by analytical model inversion. In Sec. \ref{sec:results} we apply the semi-analytical PASTIS method to the baseline APLC coronagraph design for LUVOIR-A, where we also provide further insight into the accuracy and validity of the framework, before we close with conclusions and an outlook on future work in Sec. \ref{sec:CONCLUSION}.

\section{PASTIS model of telescope segment-level aberrations in high-contrast coronagraphy}
\label{sec:top-level-pastis}

In this section, we first recall how the PASTIS model can be established using an analytical approach\cite{leboulleux2018}, which can be applied to both perfect and real coronagraphs. We then introduce a new semi-analytical derivation of the PASTIS model, which is more readily applied and suited for the analysis of actual coronagraph and telescope designs. In subsequent sections, we illustrate the application of the semi-analytical PASTIS approach to the LUVOIR-A coronagraph.

\subsection{Introduction to the PASTIS propagation model}
\label{subsec:pastis-introductoin}
\subsubsection{Analytical model derivation with a perfect coronagraph}
\label{subsubsec:perfect-coro}

\begin{figure}
   \begin{center}
   \begin{tabular}{c}
   \includegraphics[width =\textwidth]{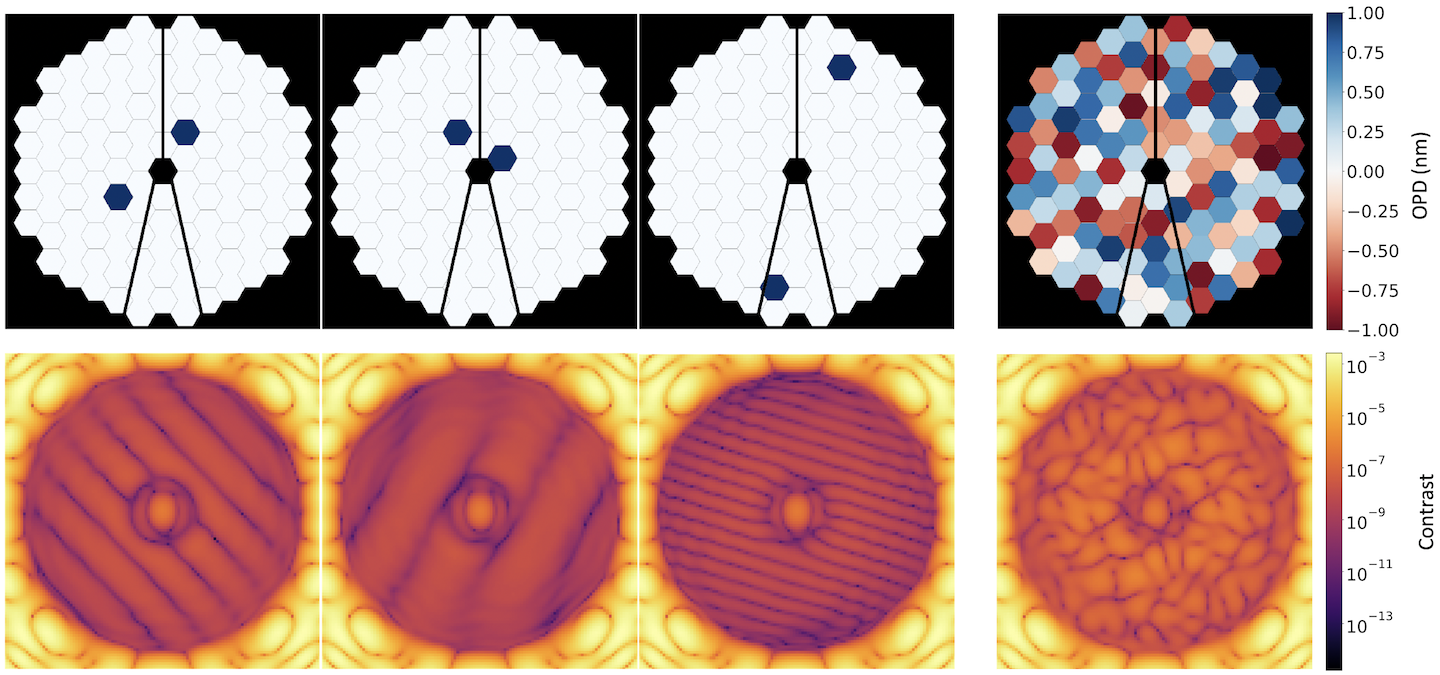}
   \end{tabular}
   \end{center}
   \caption[Piston Pairs] 
   { \label{fig:piston-pairs} 
Piston pair aberrations on a segmented pupil (top) and the resulting image plane intensity distributions in the dark-hole (bottom), using the baseline APLC for LUVOIR (see Sec.~\ref{sec:results}). The left three panels show different interference pairs with corresponding Young-like interference fringes, while the right panel shows a random distribution of local piston on all segments of the pupil and the resulting image plane intensities. All plots appear on the same scale.
}
\end{figure}

PASTIS approaches the problem by first formulating a model of coronagraphic images in the presence of primary mirror segment aberrations. The segments surface figure and their alignment state can be described by segment-level aberrations on the primary mirror as Zernike polynomials. Segment-level piston, tip/tilt, focus and astigmatism will be the most common or dominant aberrations for a segmented primary, for example in a three mirror anastigmat design such as used for JWST\cite{acton2004, acton2012wfsc-for-jwst, knight2012}. For PASTIS, we therefore expand the phase aberration in the pupil as a sum of local (segment-level) Zernike polynomials\cite[Eq.~9]{leboulleux2018}:

\begin{equation}
    \phi (\mathbf{r}) = \sum_{(k,l)=(1,1)}^{(n_{seg}, n_{zer})} a_{k,l}\ Z_l(\boldsymbol{\mathbf{r}}-\mathbf{r_k}),
    \label{eq:local-aberrations}
\end{equation}
where $\mathbf{r}$ is the pupil plane coordinate, $\phi$ the phase and $n_{seg}$ is the total number of segments, indexed by $k$. $a_{k,l}$ is the Zernike coefficient with Noll index\cite{noll1976} $l$ up to the maximum Zernike $n_{zer}$ and $Z_l$ is the $l^{th}$ Zernike. 
In this paper, we limit the study to a single Zernike mode (piston; index $l=0$) as illustrated in Fig.~\ref{fig:piston-pairs}. Hence we drop the $l$ index in all consecutive equations, but the PASTIS methodology is applicable to any Zernike mode or combination thereof.

High-contrast coronagraphy requires exquisite wavefront quality and therefore the PASTIS model assumes the small aberration regime where the electric field is well approximated as an affine function of the phase: 
$E(\mathbf{r}) = P(\mathbf{r})\,e^{i\phi(\mathbf{r})}\simeq P(\mathbf{r}) + i\,\phi(\mathbf{r})$, where the phase $\phi(\mathbf{r})$ is defined over the same support as the pupil aperture $P(\mathbf{r})$. For a perfect coronagraph that totally cancels all on-axis light in a theoretical Lyot plane, the overall coronagraph propagation reduces to a single Fourier transform of the linearized phase term and the final intensity is simply:
\begin{equation}
    I (\mathbf{s}) =  \left | \hat{\phi}(\mathbf{s})\right |^2,
    \label{eq:contrast_perfect_coron}
\end{equation}
with $\ \hat{ }\ $ representing the Fourier transform and omitting the scaling factors for readability; $\mathbf{s}$ represents the image plane coordinates. Using the Zernike decomposition of the phase above, we obtain:
\begin{equation}
    I (\mathbf{s}) = \left | \hat{Z} (\mathbf{s}) \sum_{k=1}^{n_{seg}} a_k e^{-i\mathbf{r_k}\cdot\mathbf{s}}\right |^2, 
    \label{eq:square-modulus}
\end{equation}

Following the derivation in the original paper, it appears that Eq.~\ref{eq:square-modulus} can be expressed as a sum of interference patterns between all segment pairs, weighed by an envelope given by the Fourier transform of the Zernike polynomial being studied (segment piston here) \cite[Eq.~10 -- 12]{leboulleux2018}:
\begin{equation}
    I (\mathbf{s}) = \left | \hat{Z} (\mathbf{s}) \right |^2 \, \sum_{i=1}^{n_{seg}} \sum_{j=1}^{n_{seg}} a_i\ a_j\ \cos[(\mathbf{r_j}-\mathbf{r_i}) \cdot \boldsymbol{s}] .
    \label{eq:interference-plus-envelope}
\end{equation}
 The coefficients $a_i$ and $a_j$ are the respective Zernike coefficients on segments $i$ and $j$. 
This expression is very similar to Young fringes for pairs of segments as shown in Fig.~\ref{fig:piston-pairs}.  
This analytical formulation of the fringes with a perfect coronagraph implicitly assumes that all segments have the same impact on the final image. This assumptions breaks for a real coronagraph and telescope design with apodizers, Lyot stops masks, or central obstruction with support structures \cite{ndiaye2016,Ruane2018jatis,St.Laurent2018spie,Belikov2018spie,Riggs2018spie,crill2019aas}. In this case the analytical model must be calibrated using the contrast from an end-to-end (E2E) simulation where an equal amplitude Zernike mode is applied to a given segment. This additional calibration step has been demonstrated in the case of an APLC using the analytical model\cite[Eq.~18, Fig.~8]{leboulleux2018}. However, this additional step is somewhat cumbersome and an approximation that can be alleviated by using numerical images to build the PASTIS model in a semi-analytical approach (Sec. \ref{subsec:semi-analytic-extension}).

\subsubsection{PASTIS generalization from perfect to real coronagraph, and matrix formalism }
\label{subsec:matrix-model}

In the more realistic situation of an non-perfect coronagraph, the model needs to account for the actual propagation of the linearized pupil electric field through the coronagraph. We assume that the coronagraph propagation can be represented by a linear operator $\mathcal{C}$, which for example is a valid assumption for an APLC. The contrast expression from Eq.~\ref{eq:contrast_perfect_coron} therefore becomes: 
\begin{equation}
  I (\mathbf{s}) = \left | \mathcal{C}\{P\}(\mathbf{s}) + i\, \mathcal{C}\{\phi\}(\mathbf{s})\right |^2  
  \label{eq:real-coro}
\end{equation}{}
This intensity is therefore the sum of three terms\cite[Eq.~16]{leboulleux2018}: a contrast floor corresponding to the coronagraphic contrast in the absence of optical aberration $\left| \mathcal{C}\{P\}(\mathbf{s}) \right|^2$, and the quadratic form  $\left| \mathcal{C}\{\phi\}(\mathbf{s})\right |^2 $ that generalizes Eq.~\ref{eq:interference-plus-envelope} where the cosine terms are no longer valid since the simple Fourier transform is replaced by the true coronagraph propagation operator $\mathcal{C}$. The spatial average of the cross-term $2 \Re\{\mathcal{C}\{P\}(\mathbf{s}) \mathcal{C}\{\phi\}(\mathbf{s})^*\}$ over a symmetrical dark-hole is zero\cite[Appendix A]{leboulleux2018}. Therefore, the average contrast can be expressed in a matrix formalism: 
\begin{equation}
    c = c_0 + \mathbf{a}^T M \mathbf{a},
    \label{eq:pastis-equation}
\end{equation}
where $c$ is the mean contrast in the dark-hole, $c_0$ the coronagraph floor (i.e. the average contrast in the dark-hole in the absence of aberrations), $M$ is the PASTIS matrix with elements $m_{ij}$, $\mathbf{a}$ is the aberration vector of the local Zernike coefficients on all $n_{seg}$ segments and $\mathbf{a}^T$ its transpose.

The PASTIS matrix $M$ can be calculated using either the analytical approach \cite[Eq.~20]{leboulleux2018}, or directly using an end-to-end simulation in the semi-analytical approach introduced in the next section. Once the PASTIS matrix has been established, we can calculate the mean dark-hole contrast corresponding to any aberration vector directly, by using Eq.~\ref{eq:pastis-equation}. This is particularly efficient since this calculation only includes linear algebra and no longer requires E2E optical simulations. 
This matrix formalism for the contrast was validated for the 36-segment ATLAST telescope pupil with an APLC to yield the same contrast results like the E2E model to within an error of 3\% \cite[Fig.~7]{leboulleux2018}. This result will also be validated in Section \ref{subsec:matrix-construction-validation} with the semi-analytical matrix construction. 

\subsection{Calculation of the PASTIS matrix through analytical and semi-analytical approaches}
\label{subsec:semi-analytic-extension}
To calculate the PASTIS matrix $M$, we evaluate the contrast from aberrated pairs of segments $i,j$ with the Zernike $Z(\mathbf{r})$, represented by the phase:
\begin{equation}
    \phi(\mathbf{r}) = a_i Z(\mathbf{r} - \mathbf{r}_i) + a_j Z(\mathbf{r} - \mathbf{r}_j).
    \label{eq:pair-phase}
\end{equation}
Based on Eq.~\ref{eq:real-coro}, the average dark-hole intensity for that aberrated pair becomes: 
\begin{equation}
    \left<I_{ij}(\mathbf{s})\right> = \left<\left | \mathcal{C}\{P\}(\mathbf{s}) \right |^2 \right> + \left< \left | \mathcal{C}\{\phi\}(\mathbf{s})\right |^2 \right>,
\end{equation}
where the average contrast floor $c_0 = \left<| \mathcal{C}\{P\}(\mathbf{s})|^2\right>$ and the quadratic form $\left<\left |\mathcal{C}\{\phi\}(\mathbf{s})\right |^2 \right>= \mathbf{a}^T M \mathbf{a}$ (Eq.~\ref{eq:pastis-equation}). 

The contrast term $c_{ij} = \left \langle I_{ij}(\mathbf{s)} \right \rangle_{DH}$ for the pair of segments $i,j$ is then:
\begin{equation}
\begin{aligned}
    c_{ij} &= \left<\left | a_i \mathcal{C}\{Z(\mathbf{r} - \mathbf{r}_i)\} + a_j \mathcal{C}\{Z(\mathbf{r} - \mathbf{r}_j)\} \right |^2 \right> + c_0 \\
    &= \left<a_i^2 \left | \mathcal{C}\{Z(\mathbf{r} - \mathbf{r}_i)\} \right|^2 \right> + \left<a_j^2 \left | \mathcal{C}\{Z(\mathbf{r} - \mathbf{r}_j)\} \right |^2 \right> + \left<a_i a_j 2 \mathcal{C}\{Z(\mathbf{r} - \mathbf{r}_i)\} \mathcal{C}\{Z(\mathbf{r} - \mathbf{r}_j)\} \right> + c_0
    \label{eq:three-terms}
\end{aligned}
\end{equation}

Eq.~\ref{eq:three-terms} can be identified with the quadratic expression from Eq.~\ref{eq:real-coro} by introducing the elements $m_{ij}$ of the
$M$ matrix as:
 \begin{equation}
     c_{ij} - c_0 = a_i^2 m_{ii} + a_j^2 m_{jj} + 2 a_i a_j\ m_{ij},
     \label{eq:almost-there-off}
 \end{equation}
where we identify the diagonal terms as: 
  \begin{equation}
     m_{ii} = \frac{c_{ii} - c_0}{a_i^2},
     \label{eq:diagonal-elements}
 \end{equation}
and therefore, the off-diagonal elements of the PASTIS matrix can then be expressed as:
 \begin{equation}
     m_{ij} = \frac{c_{ij} - c_0 - c_{ii} - c_{jj}}{2 a_c}.
     \label{eq:off-diagonal-elements}
 \end{equation}
The normalization by ${1}/{a_c^2}$ defines the units of the PASTIS matrix to be contrast divided by the same units like $a_c$, which is important when matching them with the units of the pupil aberration vector in Eq.~\ref{eq:pastis-equation}. The aberration $a_c$ used for the matrix generation has to be chosen in the valid range of the PASTIS development\cite[Sec 3.2.]{leboulleux2018}, i.e. in the small phase aberrations linear regime, but large enough to beat the coronagraph floor. This corresponds to the range of quadratic phase dependency illustrated in the hockey stick curve (Fig.~\ref{fig:hockeystick}).

We note that off-diagonal elements $m_{ij}$ in the PASTIS matrix (Eq.~\ref{eq:off-diagonal-elements}) can be negative based on their definition (i.e. if the diagonal contrasts $c_{ii}$ are large compared to the contrast contribution from the segment pair $c_{ij}$). This is not an issue as the only constraint for the matrix is to be positive semi-definite to ensure positive singular values that translate directly into sensible mode tolerances, see Sec. \ref{sec:tolerancing}.

In summary, the PASTIS matrix is constructed in two steps: (1) create pair-wise aberrated images $I_{ij}$ to measure their dark-hole mean contrast $c_{ij}$ and (2) use these contrast values to compute the PASTIS matrix $M$ with Eqs. \ref{eq:diagonal-elements} and \ref{eq:off-diagonal-elements}. 
The difference between the purely analytical and the new semi-analytical PASTIS models lies in the image simulation method, respectively using analytical images and using a numerical E2E simulator.

\section{Model inversion for tolerancing and stability study}
\label{sec:tolerancing}

The PASTIS matrix $M$ and Eq.~\ref{eq:pastis-equation} give a direct analytical expression to calculate the dark-hole mean contrast resulting from any random segment-level aberration $\mathbf{a}$. This makes PASTIS particularly well suited for error budgeting analyses for example using otherwise time consuming Monte-Carlo analyses. 

Moreover, by inverting this analytical model, we can reverse the calculation and determine the pupil plane aberration vector $\mathbf{a}$ that meets a specific average contrast target. We use a singular value decomposition (SVD) of the PASTIS matrix $M$ to perform this inversion. The SVD produces a set of singular values $\lambda_p$ and singular modes $\mathbf{u}_p$, where $p$ goes from $1$ to $n_{modes}$ and $n_{modes}$ will in general be equal to $n_{seg}$ (see also Sec. \ref{subsec:modes-and-per-segment}). The modes $\mathbf{u}_p$ form an orthonormal basis set that allows us to express any arbitrary pupil plane aberration $\mathbf{a}$ as a linear combination of the modes $\mathbf{u}_p$ with weighting factors $\sigma_p$:
\begin{equation}
    \mathbf{a} = \sum_{p=1}^{n_{modes}} \sigma_p \mathbf{u}_p.
    \label{eq:lin-combo}
\end{equation}
The analysis of the eigenmodes $\mathbf{u}_p$ provides information about the critical modes of the system that can be used to place tolerances on segment cophasing and stability. This will be illustrated for a LUVOIR-A coronagraph in the following section. 

Using Eq.~\ref{eq:pastis-equation} that connects the aberration $\mathbf{a}$ to the mean contrast $c$, we can derive the mode weights $\sigma_p$ that  yield the target contrast $c$ by introducing the decomposition of Eq.~\ref{eq:lin-combo} into Eq.~\ref{eq:pastis-equation}:
\begin{equation}
    c - c_0 = \left ( \sum_{p}^{n_{modes}} \sigma_p \mathbf{u}_p \right )^T M \left ( \sum_{p}^{n_{modes}} \sigma_p \mathbf{u}_p \right ).
    \label{eq:sigma_deriv-1}
\end{equation}

Assuming that each individual mode $\mathbf{u}_p$ contributes a fraction $c_p$ of the total mean contrast $c$ (see also Sec. \ref{subsec:modes-and-per-segment}), we can express this contrast contribution as:
\begin{equation}
  \begin{aligned}
    c_p &= \left (  \sigma_p \mathbf{u}_p \right )^T M \left (  \sigma_p \mathbf{u}_p \right ) \\
    &= \sigma_p^2 \mathbf{u}_p^T \, M \,\mathbf{u}_p \\
    &= \sigma_p^2 \lambda_p.
    \label{eq:sigma_deriv-2}
  \end{aligned}
\end{equation}
This approach enables us to calculate the static mode tolerances $\sigma_p$ directly from the individual contrast contribution $c_p$ (see Sec. \ref{subsec:modes-and-per-segment}) and the eigenvalues $\lambda_p$ \cite[Eq.~29]{leboulleux2018}):
\begin{equation}
    \sigma_p = \sqrt{\frac{c_p}{\lambda_p}}.
    \label{eq:calc-sigma}
\end{equation}
This allows us to compute the static mode tolerances $\sigma_p$ for each PASTIS mode $\mathbf{u}_p$. 
We can then collapse all modes into per-segment aberration tolerances $\mu_k$ for a given target contrast: 
\begin{equation}
    \mu_k = \sum_p^{n_{modes}} u_{p,k}^2\ \sigma_p
    \label{eq:mus}
\end{equation}
 This allows to set a target contrast based on scientific requirements, and then directly determine the maximum static aberration tolerances for the system to meet the target contrast. 
 
\section{Static wavefront sensitivities for coronagraphy with LUVOIR~A}
\label{sec:results}
The LUVOIR-A coronagraphic instrument includes a suite of three APLC coronagraphs  with focal mask diameters that maximize the exo-Earth yield in both detection and characterization\cite{luvoir-interim-report}\footnote{LUVOIR final report: \url{ https://asd.gsfc.nasa.gov/luvoir/resources/docs/LUVOIR_FinalReport_2019-08-26.pdf}}.  The smallest focal plane mask (FPM) coronagraph, considered in this section, is typically used for spectroscopic characterization in the wavelength band where molecular oxygen and water can be detected ($0.76\, \mu m$ and $0.94\, \mu m$). Planet detection can however be performed at shorter wavelengths (e.g. around $0.4\, \mu m$) where a given angular size corresponds to larger inner working angle in diffraction resolution units ($\lambda/D$). With an APLC, this larger focal plane mask produces a higher throughput and more robust coronagraph design, which is where the trade-off between the three designs is made.

\begin{figure}
   \begin{center}
   \begin{tabular}{c}
   \includegraphics[width = \textwidth]{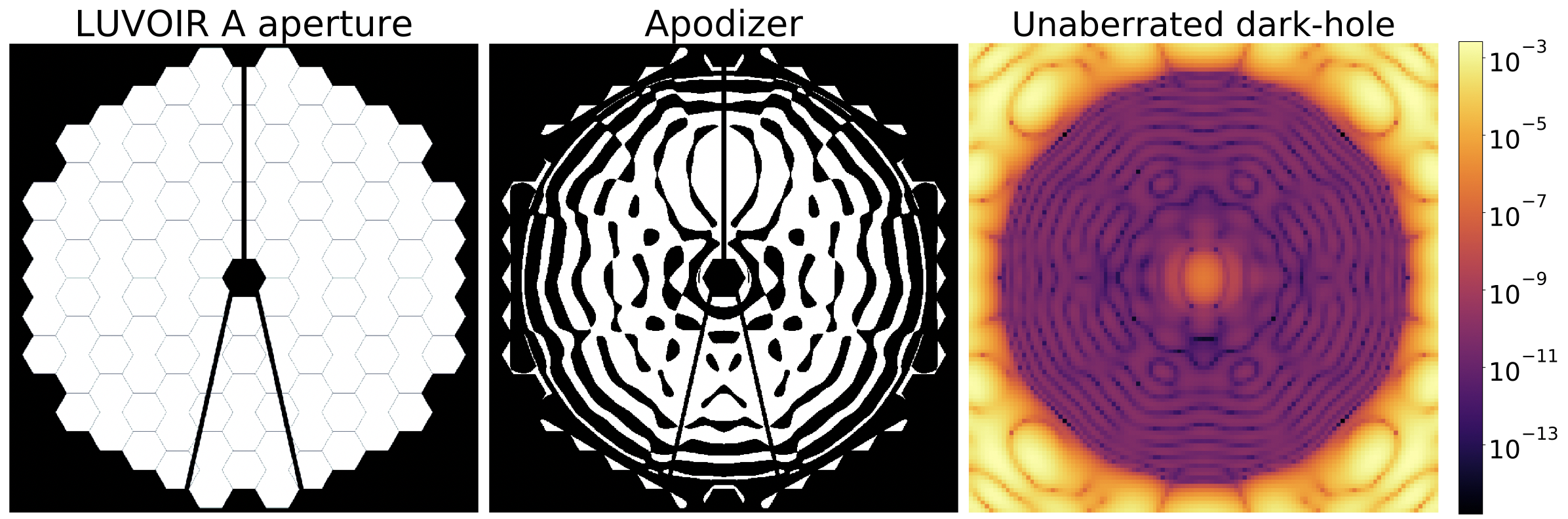}
   \end{tabular}
   \end{center}
   \caption[Aperture apod dh] 
   { \label{fig:aperture-apod-dh} 
\textit{Left:} LUVOIR-A design aperture with a diameter of 15m. \textit{Middle:} baseline apodizer for the LUVOIR-A APLC, intended for exoplanet characterization. It uses an FPM with a radius of 3.5 $\lambda/D$. \textit{Right:} Resulting coronagraphic image, with a dark-hole from 3.4 to 12 $\lambda/D$ and a mean normalized intensity of $4.2 \times 10^{-11}$, which is the coronagraph floor in the absence of optical aberrations.}
   \end{figure}

The LUVOIR-A aperture and the baseline apodizer are shown in Fig.~\ref{fig:aperture-apod-dh}. The corresponding FPM has a radius of $3.5 \lambda/D$, followed by a hard edge annular Lyot stop, which inner and outer diameters are 12.0\% and 98.2\% of the circumscribed diameter of the apodizer. 
The resulting coronagraphic image of this optical system is shown in the right of the same figure, with an average coronagraph floor of $4.2 \times 10^{-11}$ in the absence of optical aberrations.

All of the presented work was developed in Python and made publicly available in the PASTIS package\cite{pastis-software}.

\subsection{PASTIS matrix construction and validation}
\label{subsec:matrix-construction-validation}

\begin{figure}
   \begin{center}
   \begin{tabular}{c}
   \includegraphics[width =\textwidth]{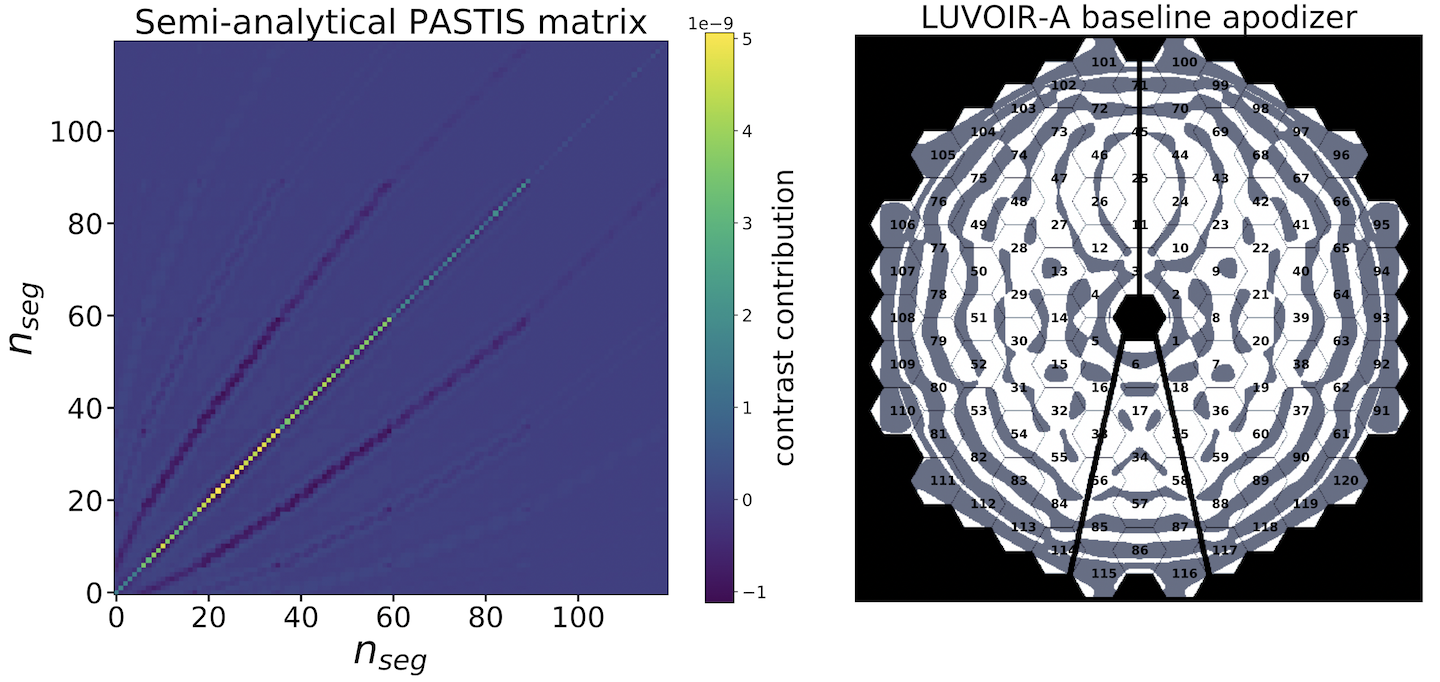}
   \end{tabular}
   \end{center}
   \caption[Numerical matrix] 
   { \label{fig:numerical_matrix} 
\textit{Left:} Semi-analytical PASTIS matrix of the 120 segment LUVOIR-A design with the baseline APLC. This matrix is symmetric by construction and the dark streaks are negative values. The diagonal elements show directly which segments have more impact on the contrast than others, e.g. the outer most ring of the telescope (segments 80-120) has lower values, indicating thee higher apodization fraction of these segments.  \textit{Right:} Apodizer overlapping with the telescope aperture. This shows how some segments are more obstructed by the apodizer than others, e.g. the outer two rings and the innermost ring have more black area than the rest of the segments, which is also reflected in the diagonal elements of the PASTIS matrix to the left.}
   \end{figure}

The semi-analytical PASTIS matrix for this coronagraph is calculated following Sec.~\ref{subsec:semi-analytic-extension} and shown in Fig.~\ref{fig:numerical_matrix}. We chose an ad hoc value of $a_c = 1$~nm in the middle of the valid range for the PASTIS model for this particular apodizer, as discussed in Sec. \ref{subsec:semi-analytic-extension}.

The PASTIS matrix shows how some segments have a higher impact on the final contrast than others. This is visible along the diagonal, which records the contrast contribution from each individual segment alone. For example, segments 61-120 have a lower contrast contribution, as they correspond to the darker areas of the apodizer on the outer two rings of the aperture (see Fig.~\ref{fig:numerical_matrix}, right panel). This effect is also visible on the innermost ring of hexagons.

We can also notice streaks of negative values in the matrix in the off-axis areas, as discussed in Sec. \ref{subsec:semi-analytic-extension}.
   
\begin{figure}
   \begin{center}
   \begin{tabular}{c}
   \includegraphics[width = 0.7\textwidth]{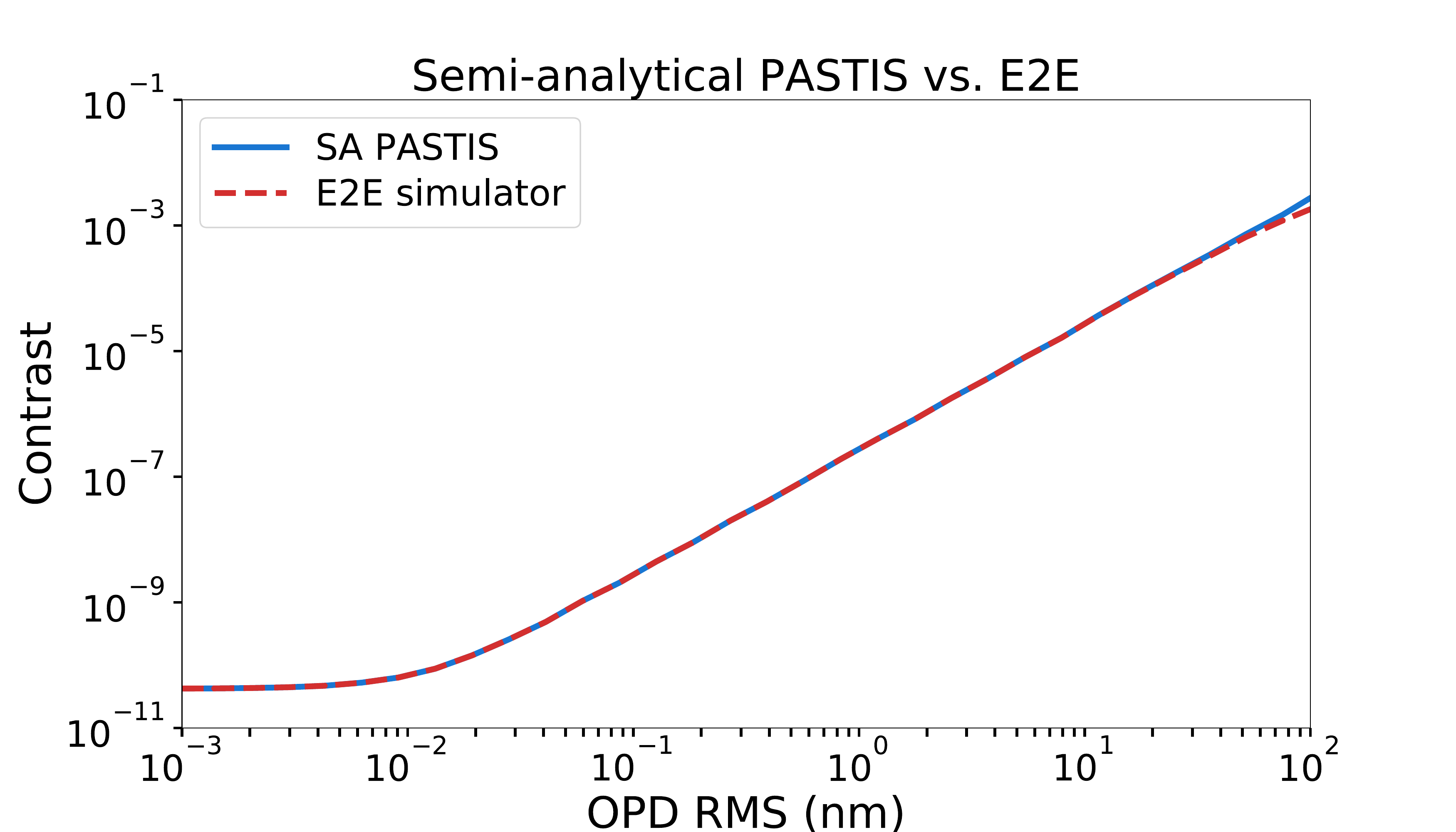}
   \end{tabular}
   \end{center}
   \caption[Contrast validation] 
   { \label{fig:hockeystick} 
Coronagraphic dark-hole contrast defined as the normalized intensity to peak of direct image in the dark-hole, and evaluated as a function of the segment phasing error (in~nm RMS). For each surface error amplitude, the contrasts are computed from both the end-to-end simulator (dashed red) and the PASTIS method (full blue). The behavior is a hockey-stick graph where the contrast is limited by the coronagraph itself at low surface errors, building the flattened out curve to the left. From about 10 to 20~pm to a few nm, the contrast is limited by the phasing aberrations. In this range the estimation error of PASTIS is $0.06 \%$. The shown curve plots the mean of 10 random realizations for each RMS value, both for the E2E simulator as well as for the PASTIS propagation.}
   \end{figure}

We validate the PASTIS matrix by comparing the results from the PASTIS contrast with Eq.~\ref{eq:pastis-equation} to those from an E2E simulator using the same inputs and show the comparison in Fig.~\ref{fig:hockeystick}. The coronagraph floor is present at $4.2 \times 10^{-11}$ and both models almost perfectly overlap with an error of $0.06 \%$. The accuracy of the semi-analytical approach is significantly higher than that of the fully analytical matrix because the construction of the PASTIS matrix is based on the actual E2E simulation as opposed to a post-calibrated analytical fringe model.

\subsection{PASTIS modes and per-segment tolerancing}
\label{subsec:modes-and-per-segment}

\begin{figure}
   \begin{center}
   \begin{tabular}{c}
   \includegraphics[width = \textwidth]{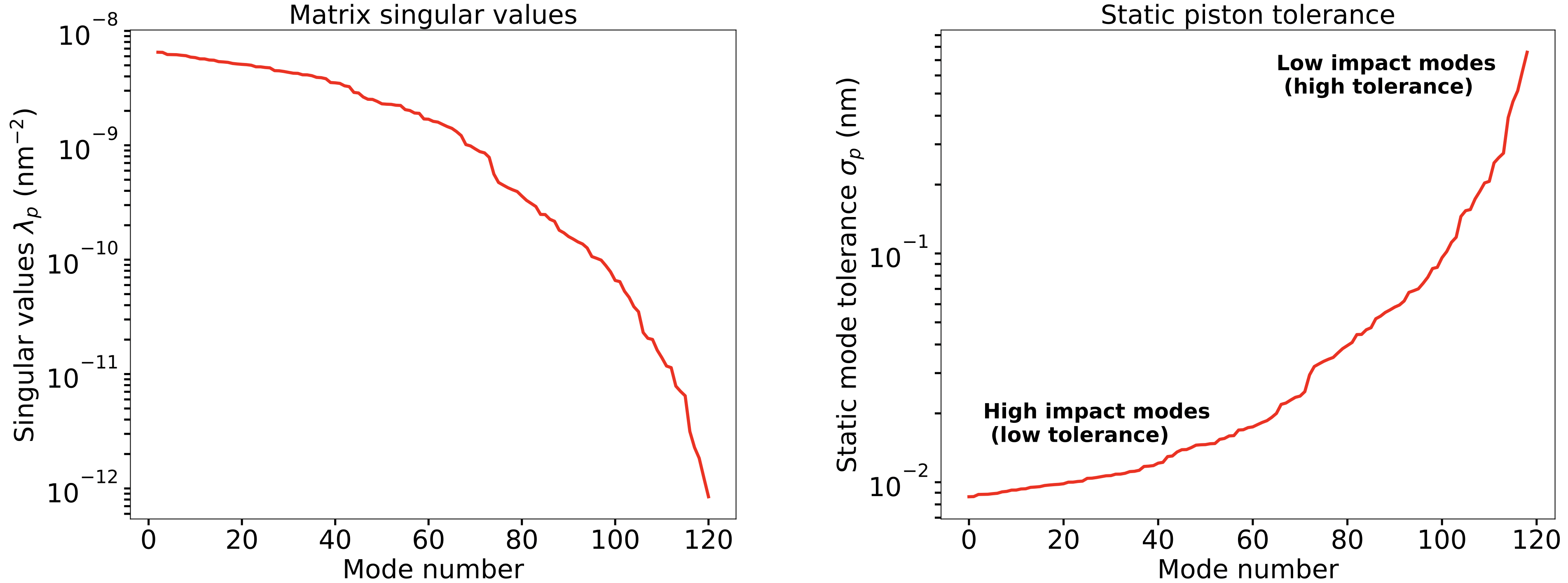}
   \end{tabular}
  \end{center}
  \caption[singular values and sigmas] 
   { \label{fig:svalues-and-sigmas} 
\textit{Left:} Singular values for the piston PASTIS matrix of the LUVOIR-A telescope with the baseline coronagraph design. Note how the PASTIS matrix does not depend on the target contrast, but it does on the choice of telescope geometry and coronagraph. \textit{Right:} Static mode tolerances for local piston aberrations for the aforementioned telescope and coronagraph with a target contrast of $c_{target} = 10^{-10}$.}
   \end{figure}

The system's singular values (see Fig.~\ref{fig:svalues-and-sigmas} left) and a set of orthonormal eigenmodes are obtained from an SVD of the PASTIS matrix. We discard the first mode since it corresponds to a global piston with singular value zero and infinite tolerance, and reduce the basis to a total number of $n_{modes} = 119$ modes. 

Using the target contrast $c_{target} = 10^{-10}$, we calculate the static segment constraints with Eq.~\ref{eq:calc-sigma}, which correspond to the maximum amplitudes of each mode $\mathbf{u}_p$ that, when all combined yield a contrast $c_{target}$. In this paper we simply allocate the same fraction of the final contrast to each mode. With $n_{modes}=119$ modes contributing to $c_{target}$ and all fractions $c_p$ being equal, $c_p = c_{target}/n_{modes}$, the static mode tolerances $\sigma_p$ become: 
\begin{equation}
    \sigma_p = \sqrt{\frac{c_{target} - C_0}{n_{modes}\ \lambda_p}}, 
    \label{eq:calc-sigma-full}
\end{equation}
which are illustrated in Fig.~\ref{fig:svalues-and-sigmas} for a target contrast of $10^{-10}$.

\begin{figure}[p]
    \centering
   \includegraphics[width = \textwidth]{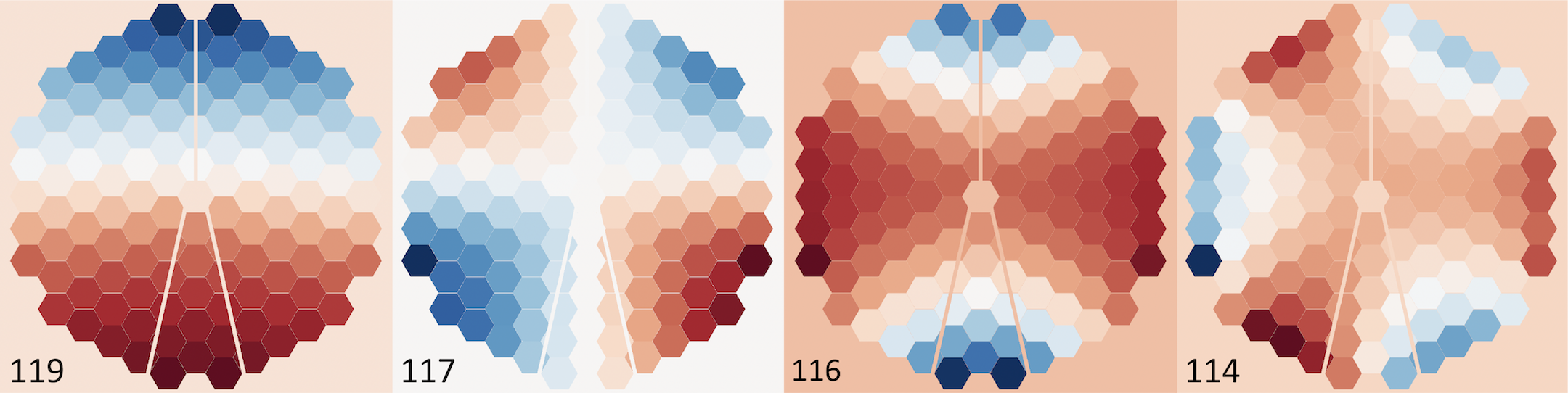}
  \caption[Low Order Modes] 
   { \label{fig:low-order-modes} 
Low-impact modes with high tolerances for the baseline APLC on the LUVOIR-A telescope, for local piston aberrations. These modes have little impact on the final contrast - they are essentially discretized Zernike modes and the coronagraph rejects them very well by design. Mode number 119 to the very left has the highest tolerance with 758~pm of piston. For comparison, the tolerances for modes 117, 116 and 114 are 543~pm, 461~pm and 274~pm, respectively.}
    \bigskip
   
   \includegraphics[width = \textwidth]{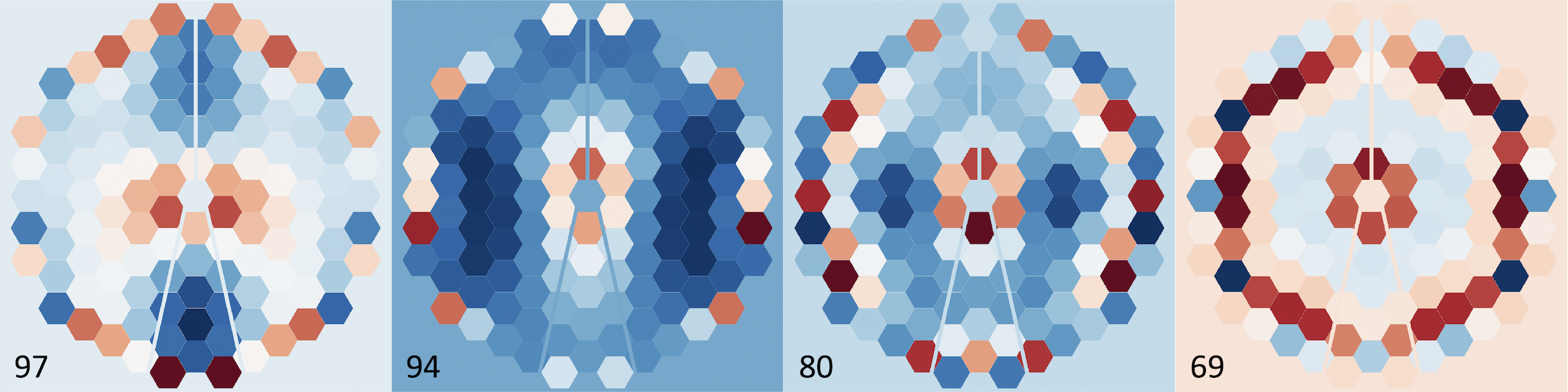}
   \caption[Mid Order Modes] 
   { \label{fig:mid-order-modes} 
Mid-impact modes with medium tolerances for the baseline APLC on the LUVOIR-A telescope, for local piston aberrations. These modes have medium impact on the final contrast, relatively speaking. These modes show mostly low-order features except for high spatial frequency components in the parts of the pupil where the apodizer covers most of the segments. For comparison, the tolerances for modes 97, 94, 80 and 69 are 74~pm, 68~pm, 38~pm and 23~pm, respectively.}
    \bigskip
   
   \includegraphics[width = \textwidth]{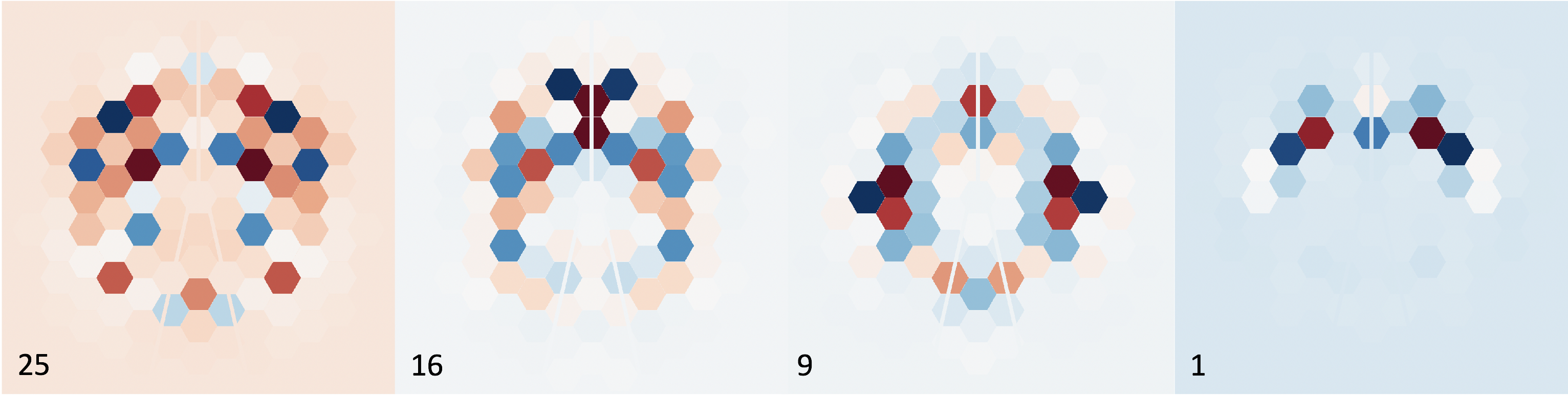}
  \caption[High Order Modes] 
   { \label{fig:high-order-modes} 
High-impact modes with low tolerances for the baseline APLC on the LUVOIR-A telescope, for local piston aberrations. These modes have the highest impact on the final contrast, with mode 1 (far right) tolerating only 8.6~pm of piston. For comparison, the tolerances for modes 25, 16, and 9 are 10.1~pm, 9.6~pm and 9.1~pm, respectively. These modes consist entirely of high spatial frequency components in the parts of the pupil where the apodizer (and other pupil plane optics) are the most transmissive.}
\end{figure}

The range of static mode tolerances $\sigma_p$ across individual modes is very large (almost two orders of magnitude from 8~pm to 758~pm). 
The modes with high mode number (100--119) have the highest tolerance per mode given the uniform contrast contribution target, and therefore the smallest sensitivity. A selection of these modes is shown in Fig.~\ref{fig:low-order-modes} and they appear as discretized low-order Zernike modes for which we know that this APLC coronagraph has a high rejection.

The modes corresponding to the mid-section of the plot are shown in Fig.~\ref{fig:mid-order-modes}. These mid-impact modes generally show low-order features, but not exclusively as they also present higher order features in the segment groups that are more concealed by the apodizer (the outer two rings and the innermost ring of the hexagonal segments), as already discussed in Sec. \ref{subsec:matrix-construction-validation} and Fig.~\ref{fig:numerical_matrix}. 

The highest-impact modes have a very low tolerance (high sensitivity) to wavefront error and are shown in Fig.~\ref{fig:high-order-modes}. These modes consist mainly of high spatial frequency components that are concentrated in sections of the pupil that are the most transmissive, i.e. where the apodizer and other pupil plane components (e.g. spiders, Lyot stop) cover the least area. The mode with the lowest tolerance is mode 1 which can only take an amplitude of 8.6~pm to remain within the target contrast range. 

\begin{figure}
   \begin{center}
   \begin{tabular}{c}
   \includegraphics[width = 0.8\textwidth]{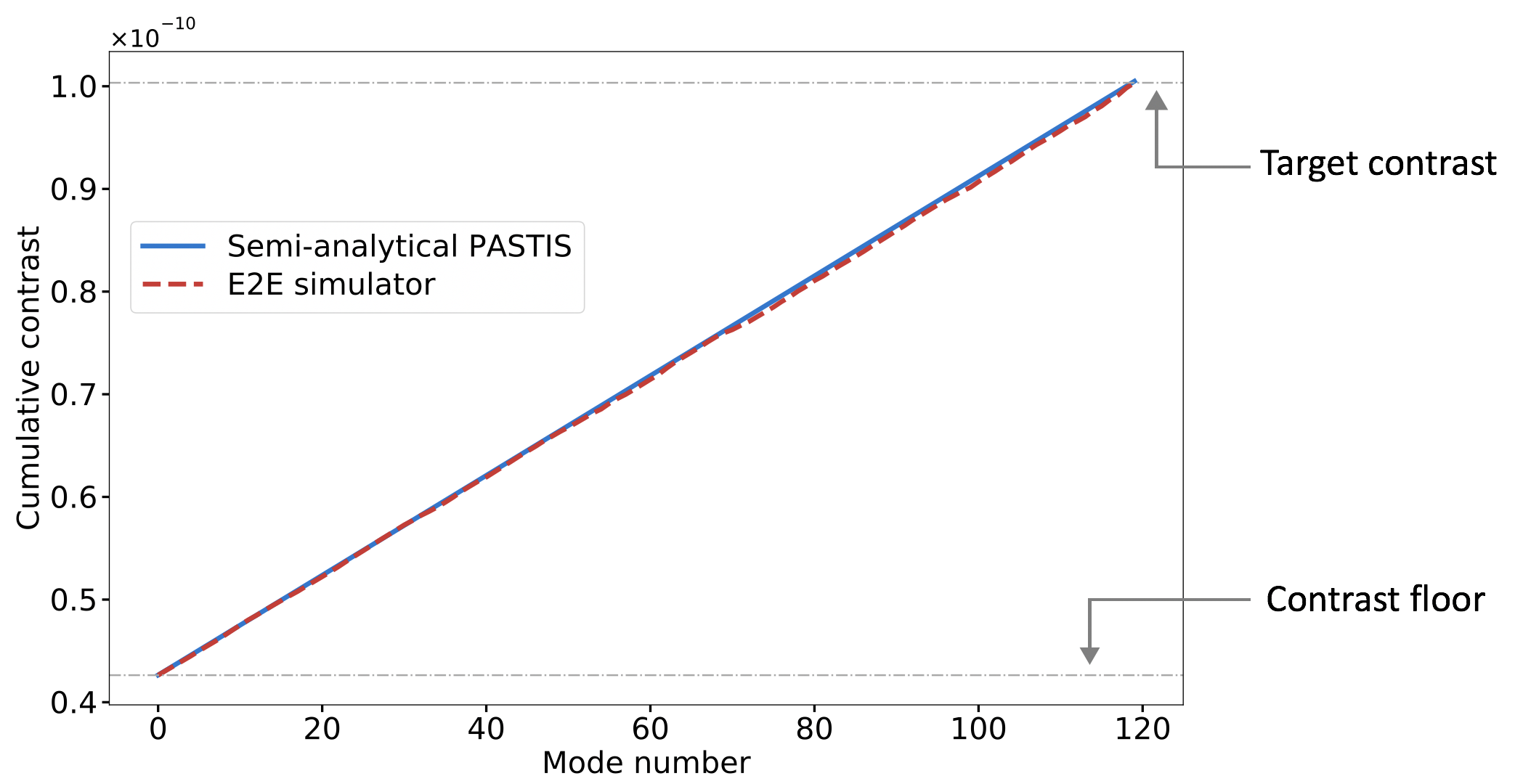}
   \end{tabular}
   \end{center}
   \caption[CumulativeContrast] 
   { \label{fig:cumulative-contrast} 
Cumulative contrast from all modes, multiplied by their respective mode tolerance $\sigma_p$, both from the PASTIS propagation and the E2E simulator. Without any aberrations applied, we get the contrast floor from the coronagraph, while application of all modes together yields the requested target contrast, here $C = 10^{-10}$. Each mode is allocated an equal contribution to the final contrast, a choice that can be revisited in more complex error budget strategies.}
   \end{figure}

We can verify that all modes contribute equally to the target contrast according to their respective mode tolerances by calculating the cumulative contrast as a function of modes weighted by their respective tolerance $\sigma_p$ (Fig.~\ref{fig:cumulative-contrast}). The first data point is at the level of the coronagraph floor (no aberrations applied), the last point hits exactly the target contrast as all modes are combined, and in between the contrast measurements are linear, as the mode contributions add up linearly in our definition of uniform mode contribution to the total contrast.

\begin{figure}
   \begin{center}
   \begin{tabular}{c}
   \includegraphics[width = \textwidth]{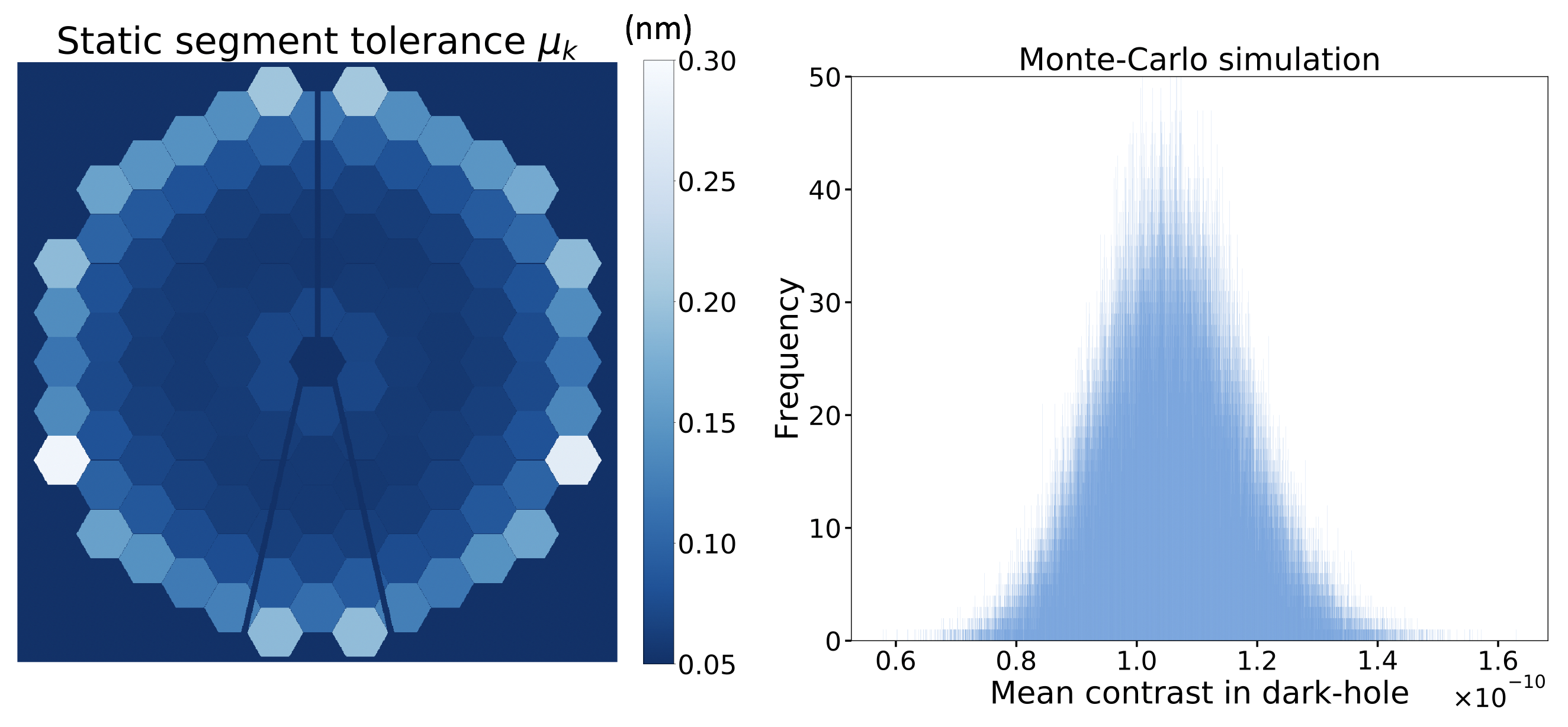}
   \end{tabular}
   \end{center}
   \caption[MonteCarlo] 
   { \label{fig:mus-and-monte-carlo} 
\textit{Left:} Static per-segment piston tolerances for the LUVOIR-A  15 m telescope with the baseline coronagraph design for local piston aberrations and a target contrast of $c_{target} = 10^{-10}$. The tolerances are not uniform across the pupil and segment groups with higher and lower impact on the final contrast are identified. \textit{Right:} Monte-Carlo E2E optical simulation of the tolerance map to the left, done for 100,000 realizations of the map weighted by a random uniform distribution between 0 and 1. Recovering the mean target contrast validates the analytical PASTIS analysis.}
   \end{figure}

PASTIS can provide an error budget for local piston aberrations on the LUVOIR-A telescope with an APLC and a target contrast of $c_{target} = 10^{-10}$ by collapsing all mode tolerances into segment-level constraints as shown in Eq.~\ref{eq:mus}. This yields a per-segment tolerance map for local piston aberrations shown in Fig.~\ref{fig:mus-and-monte-carlo}. The per-segment tolerance is not uniform over the pupil, but tracks the black and white distribution of the apodizer (see Fig.~\ref{fig:aperture-apod-dh} for reference): the highest tolerance is on a corner segment of the outermost ring, segment 110 with 290~pm while the lowest tolerance lies on segment 11 with 56~pm, located in the second ring form the center. 
This brings a significant relief for the definition of segment stability. We can directly use this information to define specific regions in the pupil for which we can relax the stability requirements, e.g. here in the outer rings which are naturally more challenging for mechanical and thermal stability. Also, the results of this tolerancing analysis could potentially be included in future coronagraph design processes to render the coronagraph robust to certain modes. This would bring new levels of trade-off between coronagraph design and telescope-level engineering constraints, helping to define parts of the observatory that have more stringent stability requirements than others.
   
To confirm the resulting numbers from the tolerance map, we run a Monte-Carlo end-to-end optical simulation where we randomize the segment pistons using a uniform distribution between 0 and $\mu_k$. The result for 100,000 samples (Fig.~\ref{fig:mus-and-monte-carlo}) shows that we can recover the mean target contrast for which the segment-level tolerance analysis was established, thus validating the analytical development for obtaining constraints on segment-based aberrations.

While this analysis only presents results for a single local Zernike (piston), previous work on PASTIS has shown that the qualitative sensitivity of the segments remains consistent for other Zernike modes and was illustrated for astigmatism\cite{leboulleux2018}. For a comprehensive quantitative analysis, a similar process to the one presented for piston will need to be repeated with an extended range of Zernikes and ultimately combinations of local aberrations on the segments.

\section{SUMMARY AND CONCLUSION}
\label{sec:CONCLUSION}

The goal of the PASTIS model for segmented aperture coronagraphy is to provide a direct and simple analytical expression of the mean dark zone contrast, as a function of segment-level aberrations. The wavefront perturbations are expressed as segment-level Zernike polynomials (in this paper, piston, but also valid for any Zernike or combination thereof). The PASTIS model is unique to the given telescope geometry,  coronagraph design, and to a segment-level modal basis of choice.

In this paper we extend the original PASTIS propagation model from a purely analytical to a semi-analytical method, in which we substitute the use of analytical images with numerically simulated images, in order to build the PASTIS matrix. The PASTIS evaluation of the mean dark zone contrast is orders of magnitude faster than with classical E2E simulators, and with the semi-analytical approach, it is more accurate than the fully analytical solution.

This analytical propagation model based on the PASTIS matrix can be inverted, which permits the derivation of an error budget for segment-level cophasing errors, depending solely on the target contrast for science observations. A singular value decomposition of the PASTIS matrix yields a set of orthonormal PASTIS modes that will influence the image plane mean contrast depending on their derived mode-level tolerances $\sigma_p$. Assuming a uniform contrast contribution by each mode (a choice that can be adjusted to any other error budget strategy) these maximum mode contributions range from 8~pm to 758~pm for the baseline small coronagraph design of the LUVOIR-A telescope, with a target contrast of $C = 10^{-10}$. The same mode-based tolerances can be collapsed into a segment-level tolerance map that shows a minimum and maximum wavefront error constraint between 56~pm and 290~pm per segment for the same setup. We observe how this provides a local relaxation of the wavefront error limits on certain parts of the pupil, which can be exploited for example for the backplane mechanical design and observatory-level control strategy. 

The semi-analytical PASTIS approach is therefore a flexible tolerancing tool that can be adapted readily to any telescope geometry or coronagraph. This enables us to perform active trade-offs between coronagraph designs that will provide certain modal rejections and telescope-level engineering constraints, implemented in other parts of the observatory. 

The analysis presented in this paper is purely static; however, the extension to dynamical drift rates also depends on the observing scenario and wavefront control strategy, which will put this propagation model on different time scales\cite{pueyo2019aas}. 

Future work will address such dynamic analysis methods for continuous wavefront sensing and control cases. Moreover, we need to extend the aberration basis for PASTIS applications: first to other individual Zernike modes (tip/tilt, focus, astigmatism, etc.) and then to their arbitrary combinations. The feasibility of this has already been shown in the analytical approach\cite{leboulleux2018} and should hence be regarded as a mere functional addition. Further contribution to the understanding of segment-level cophasing errors and stability will be provided through the analysis of scaling laws that will address the sensitivity to system parameters like segment number, size and shape, their relative size with respect to the total telescope pupil, geometrical arrangement, or different coronagraph types. Finally, by combining the knowledge emerging from our studies with established end-to-end simulation results will provide deeper comprehension of wavefront error tolerancing on segmented telescopes.

\acknowledgments 
This work was supported in part by Ball Aerospace and Technologies Corporation subcontract No.18KMB00077  (PI: R. Soummer, Sci-PI: L. Pueyo) as part of the Ultra-Stable Telescope Research and Analysis (ULTRA) Program funded by NASA ROSES 2017 D.15. The work was also funded by the Jet Propulsion Laboratory subcontract No.1539872 (Segmented-Aperture Coronagraph Design and Analysis; PI: R. Soummer), and the STScI Director's Discretionary Research Funds. It is also partly funded by the French national aerospace research center ONERA (Office National d'\'{E}tudes et de Recherches A\'{e}rospatiales) and by the Laboratoire d'Astrophysique de Marseille (LAM).

This research was developed in Python\footnote{\url{https://www.python.org}}, an open source programming language, and made use of the Numpy\cite{numpy1, numpy2}, Matplotlib\cite{matplotlib, matplotlib-zenodo} and Pandas\cite{pandas} packages. This research made use of PASTIS, an open-source Python package for segment-level error budgeting of segmented telescopes\cite{pastis-software}. This research made use of HCIPy, an open-source object-oriented framework written in Python for performing end-to-end simulations of high-contrast imaging instruments\cite{hcipy}. This research made use of Astropy,\footnote{\url{http://www.astropy.org}} a community-developed core Python package for Astronomy \cite{astropy:2013, astropy:2018}. I. Laginja is also thankful to Anand Sivaramakrishnan for help and discussions.

\bibliography{bibliotheque}
\bibliographystyle{spiebib}

\end{document}